\documentclass[3p,times,twocolumn]{elsarticle}
 \biboptions{comma,sort&compress}
 
\usepackage{graphicx}
\usepackage{here}
\usepackage{ecrc}

\usepackage{amssymb}

\usepackage{amsmath}

\def\lQ{\Lambda_{\rm QCD}}

\def\j{{\cal J}}

\def\m{{\cal M}}
\newcommand{\be}{\begin{equation}}
\newcommand{\ee}{\end{equation}}
\newcommand{\bea}{\begin{eqnarray}}
\newcommand{\eea}{\end{eqnarray}}
\newcommand{\nn}{\nonumber}

\volume{00}

\firstpage{1}

\journalname{Nuclear and Particle Physics Proceedings}

\runauth{}

\jid{nppp}

\jnltitlelogo{Nuclear and Particle Physics Proceedings}

\usepackage{amssymb}
\usepackage[figuresright]{rotating}

\begin{document}

\begin{frontmatter}

%%
%%%%%%%%%%%%%%%%%%%%%%%%%%%%%%%%%%%%%%%%%%%%%%%%%
\title{ Heavy Quarkonium Hybrids
 $^*$}
 % \corref{cor0}}
 \cortext[cor0]{Talk given at 20th International Conference in Quantum Chromodynamics (QCD 17),  3 - 7 july 2017, Montpellier - FR}
 \author[label1]{Joan Soto}
%  \cortext[cor0]{FAPESP CNPq-Brasil PhD student fellow.}
\ead{joan.soto@ub.edu}
\address[label1]{
Universitat de Barcelona, Departament de F\'\i sica Qu\`antica i Astrof\'\i sica i
Institut de Ci\`encies del Cosmos,  Facultat de F\'\i sica, Mart\'\i $\,$ i Franqu\`es 1, 08028 Barcelona, Catalonia, Spain}

\pagestyle{myheadings}
\markright{ }
\begin{abstract}
We report on a recent investigation on heavy quarkonium hybrids that goes beyond the usual Born-Oppenheimer approximation by including not only the mixing between nearby hybrid states but also the mixing with quarkonium states. 
We use a systematic effective field theory framework based on NRQCD together with lattice QCD inputs. Short and long distance constraints from weak coupling pNRQCD and the QCD effective string theory are also employed. 
We calculate the quarkonium and hybrid spectrum for charmonium and bottomonium, and estimate a number of decay widths. Most of the isospin zero $XYZ$ resonances fit in our spectrum either as quarkonia or as hybrid states.
 The mixing of hybrid states with quarkonia produces enhanced spin symmetry violations, which are instrumental to understand certain decays. We also present new results on the hyperfine splittings. 
\end{abstract}
% \begin{document}
\begin{keyword}  
%% keywords here, in the form: keyword \sep keyword
Heavy Quarkonium, Hybrids, NRQCD, pNRQCD, QCD string
%% MSC codes here, in the form: \MSC code \sep code
%% or \MSC[2008] code \sep code (2000 is the default)

\end{keyword}

\end{frontmatter}
%%%%%%%%%%%%
%\vspace*{-1.5cm}
\section{Introduction}

Exotic hadrons, namely those beyond quark-antiquark or three quark (antiquark) states, have been contemplated as a theoretical possibility since the early days of QCD  \cite{Jaffe:1975fd}. The interest
on exotic hadrons has recently experienced a revival due to the pletora of charmonium, and some bottomonium, resonances, the so called $XYZ$ states, discovered in the last decade that do not easily fit in the quark model spectrum (see \cite{Olsen:2017bmm,Chen:2016qju} for recent reviews). The fact that the charm and bottom quark masses ($m_c$ and $m_b$) are much larger than the typical hadronic scale, $\lQ$, has allowed to unambiguously identify tetraquark \cite{Choi:2007wga,Aaij:2014jqa,Belle:2011aa} and pentaquark \cite{Aaij:2015tga} states. We shall focus here on heavy charmonium and bottomonium hybrids, namely $c\bar c$ and $b\bar b$ states with a non-trivial gluon content, with the aim to understand at least part of the spectrum of isospin zero $XYZ$ states.

Since $m_c$, $m_b \gg \lQ$, the heavy quarks move slowly so that they see an instantaneous potential as the effective interaction. In early string models the hybrid potential was associated with the excitations of the string \cite{Giles:1977mp,Horn:1977rq}. The use of the Born-Oppenheimer (BO) approximation to obtain the hybrid potential was initiated in \cite{Hasenfratz:1980jv} within the bag model approach. The first lattice QCD calculation calculation of the hybrid potential was carried out in \cite{Griffiths:1983ah}. More recently, the BO approximation has been revisited in relation with the $XYZ$ states \cite{Braaten:2014qka}. It has also been incorporated into an effective field theory framework in \cite{Berwein:2015vca} (see also \cite{Brambilla:2017uyf}) elaborating on the weak coupling regime of pNRQCD \cite{Pineda:1997bj,Brambilla:1999xf}, and in \cite{Oncala:2016wlm,Oncala:2017hop} elaborating on the strong coupling regime of pNRQCD \cite{Brambilla:1999xf,Brambilla:2000gk,Pineda:2000sz}. The following sections are based on ref. \cite{Oncala:2017hop}, except for the section on the hyperfine splitting, which presents new material \cite{Pere}. 

\section{Quarkonium}

In order to set the scale of the hybrid spectrum it is important to have the quarkonium spectrum calculated in the same framework. Since lattice calculations exist for both the quarkonium and the hybrid potentials (see Fig. \ref{kuti}), we shall fix the single arbitrary constant of all these potentials by fitting to the charmonium and bottomonium spectrum. The shape of the quarkonium potential evaluated on the lattice ($\Sigma_g^+$ in Fig. \ref{kuti}) is well described by the Cornell potential, which has the short and long distance behavior expected from QCD perturbation theory and the QCD effective string theory (EST) \cite{Luscher:2002qv,Luscher:2004ib} respectively,

\begin{equation}
\label{V_g}
V_{\Sigma_{g}^{+}}(r)\approx -\frac{\sigma_g}{r}+\kappa_g r+E_g^{Q\bar{Q}} \, ,
\end{equation} 
where we take,
\be
\label{st}
\sigma_g=0.489 \, ,\hspace{1cm} \kappa_g=0.187\, {\rm GeV}^2 \, ,
\ee 
and we obtain from the comparison with the experimental spectrum,
\be	
	E_g^{c\bar{c}}=%2m_c
	-0.242 \, {\rm GeV} \hspace{1cm} E_g^{b\bar{b}}=%2m_b
	-0.228\, {\rm GeV} \, .
	\ee
	Note that $E_g^{Q\bar{Q}}$ should be flavor independent, and indeed $E_g^{c\bar{c}}$ and $ E_g^{b\bar{b}}$ agree within a $6\%$. The spectrum obtained with the potential above is displayed in Tables \ref{cEspectrum} and 
 \ref{bEspectrum}.%of \cite{Oncala:2017hop}.
	
	\section{Hybrids}
	
	\vspace{-1cm}
	\begin{figure}[htb]
	\centering
\includegraphics[width=8cm,height=10cm]{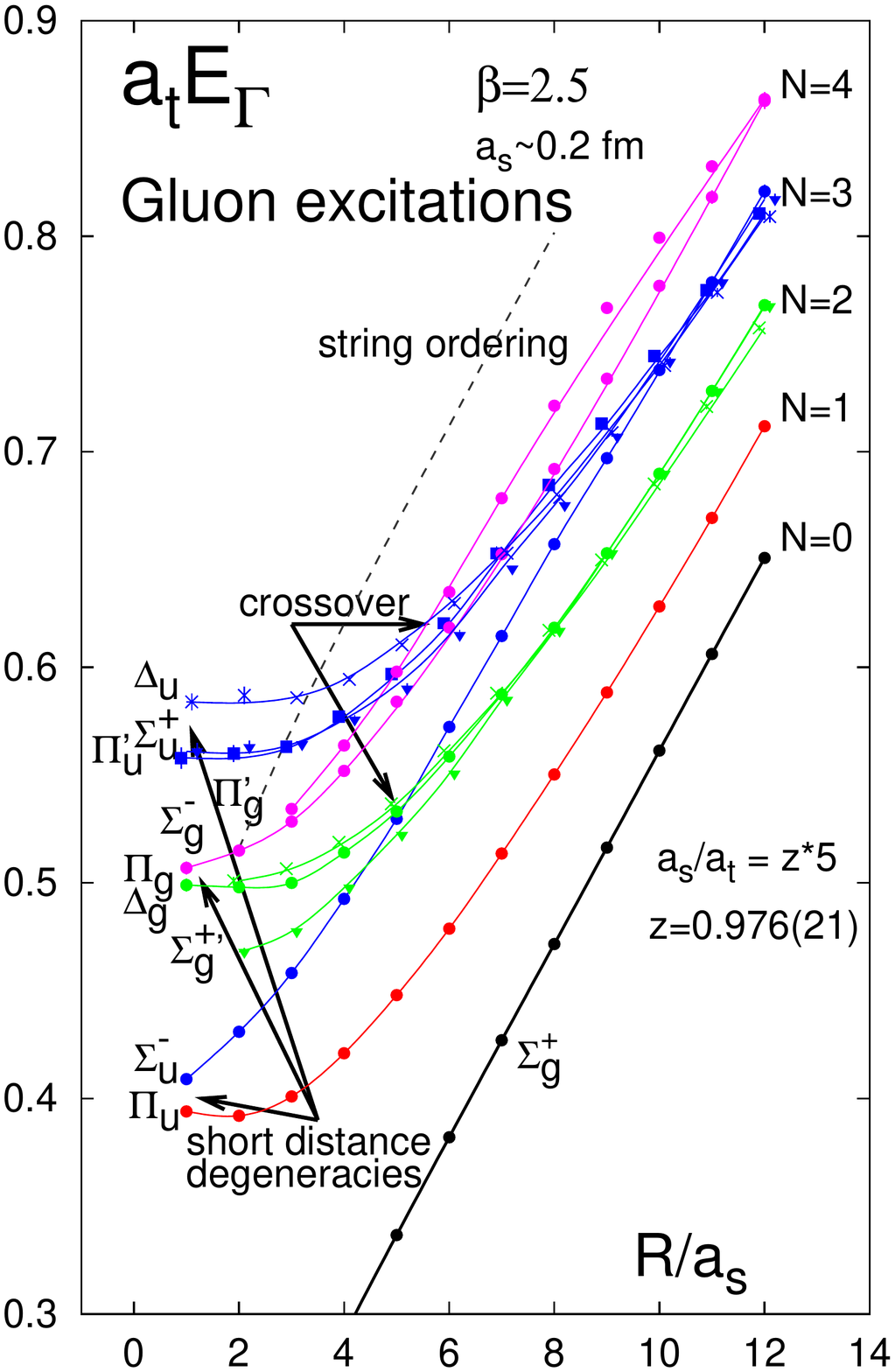}
\caption{Energy spectrum in the static limit for $n_f=0$ \cite{Juge:2002br}.}
\label{kuti}       
\end{figure}
	
	The hybrid potentials together with the quarkonium potential ($\Sigma_g^+$) are displayed in Fig. \ref{kuti}. The labels correspond to the representations of the $D_{\infty h}$ group, the group of a diatomic molecule. At short distances all the hybrid potentials must approach the repulsive Coulomb potential of the color octet configuration, as perturbation theory dictates. Furthermore, the states should gather in short distance multiplets
	according to the rotational group \cite{Brambilla:1999xf}. At long distance they must approach the behavior dictated by the QCD EST, namely to the same linear potential as the quarkonium case ($\Sigma_g^+$) with a subleading $1/r$ behavior that depends on the string state
\cite{Luscher:2004ib}. Notice that the hybrid potentials, unlike the quarkonium one, have a classical minimum, which must sit at $r\sim 1/\lQ$ (there is no other scale available). Hence the small energy fluctuations about this minimum have a size $\sqrt{\lQ^3/m_Q}$, which is parametrically smaller than $\lQ$. Hence, if we are only interested in the lower lying states for each potential, we are in a situation similar to the strong coupling regime of pNRQCD \cite{Brambilla:1999xf}, in which the scale $\lQ$ can be integrated out. This means that, in a leading approximation, we can ignore the interaction with any other hybrid or quarkonium state with an energy $\gtrsim \lQ$ above or below the low lying states. In Fig. \ref{kuti},  we observe that the short distance degeneracies are already noticeable close to the minima. Hence, it is natural to chose the degrees of freedom of the effective theory as a wave function field that describe the corresponding short distance multiplet. We are going to restrict ourselves to the lower lying hybrid multiplet, namely that formed by $(\Sigma_u^-, \Pi_u)$. At short distances, this wave function field corresponds to a quark-antiquark in a color octet state together with a chromomagnetic field that makes the whole operator color singlet \cite{Brambilla:1999xf}. Hence, we choose a vectorial wave function matrix ${\bf H}({\bf 0},{\bf r}, t)$ with the same symmetry transformations as that operator. Namely, it transforms as ${\bf H} \to h_1{\bf H}h_2^\dagger$, $h_1$, $h_2 \in SU(2)$ under spin symmetry
and as follows under parity, time reversal and charge conjugation,
\bea
\label{symH}
P: & \, {\bf H}({\bf R},{\bf r}, t)& \rightarrow -{\bf H}(-{\bf R},-{\bf r}, t)\nn\\
T: &\, {\bf H}({\bf R},{\bf r}, t)& \rightarrow -\sigma^2{\bf H}({\bf R},{\bf r}, -t)\sigma^2\\
C: &\,  {\bf H}({\bf R},{\bf r}, t)& \rightarrow -\sigma^2{\bf H}^T({\bf R},-{\bf r}, t)\sigma^2\,,\nn
\eea
As a consequence, the $P$ and $C$ quantum numbers of a hybrid state with quark-antiquark orbital angular momentum $L$ and quark-antiquark spin $S$
are,
\be
P=(-1)^{L+1} \; ,\, C=(-1)^{L+S+1} \,.
\ee
The Hamiltonian at leading order (BO approximation) is chosen 
	such that the projection of ${\bf H} $ to {\bf r} 
evolves with $V_{\Sigma_u^-}$ 
and the projection orthogonal to {\bf r} with $V_{\Pi_u}$ in the static limit. That is,
\bea
\label{h}
\!\!\!\!\!\!\!\!\!\!\!\!\!\!\!\!&{\cal L}={\rm tr} \left( {H^i}^\dagger\left( \delta_{ij}i\partial_0-{h_H}_{ij}\right)H_j \right)&\\
\!\!\!\!\!\!\!\!\!\!\!\!\!\!\!\!&{h_{H}}_{ij}\!=\!\left(\!-\frac{\nabla^2}{m_Q}\!+\!V_{\Sigma_u^-}(r)\right)\delta_{ij}\!+\!\left(\delta_{ij}\!-\!\hat{r}_{i} \hat{r}_{j}\right)\left[V_{\Pi_u}(r)\!-\!V_{\Sigma_u^-}(r)\right]&.\nonumber
\eea
In addition to  $L$ and $S$ defined above, we characterize the states with $J$, the total angular momentum of the gluons plus the orbital angular momentum of the quark-antiquark,  $\j$,  the total angular momentum of the system, and $\m$, its third component.

\subsection{Spectrum}

Spin symmetry implies that states with the same $J$ and $L$ are degenerate. They form quadruplets except for the case $J=0$ that they form a doublet. In fact, $L$ is not a good quantum number
because the potential in (\ref{h}) is not central. This leads to  coupled eigenvalue equations for $L=J\pm 1$, whereas a single (uncoupled) eigenvalue equation remains for $L=J$. We use $NL_J$ ($N=1,2,\dots$, $L=s,p,d,\dots$, $J=1,2,\dots$) to label the states in the spin symmetric limit, where $N$ is the principal quantum number. The results for the charmonium and bottomnium spectrum, both for hybrids and quarkonium, are displayed in the Tables \ref{cEspectrum} and %II of ref. \cite{Oncala:2017hop}
\ref{bEspectrum} respectively.

\begin{table}[htbp]
	\centering
	\begin{tabular}{|c|c|c|c|c|}
		\hline
		&                        &      $S=0$      &      $S=1$      &                             \\
		$NL_J$  & $M$%_{c\bar{c}}$ & $M_{c\bar{c}g}$ 
		& $\mathcal{J}^{PC}$ & $\mathcal{J}^{PC}$ & $\Lambda^{\epsilon}_{\eta}$ \\ \hline \hline
		$1s$ &  3068 &   $0^{-+}$ & $1^{--}$ & $\Sigma_g^+$ \\ 
		$2s$ &  3678 &   $0^{-+}$ & $1^{--}$ & $\Sigma_g^+$ \\ 
		$3s$ &  4131 &   $0^{-+}$ & $1^{--}$ & $\Sigma_g^+$ \\ 
		$1p_0$ &    4486 & $0^{++}$ & $1^{+-}$ & $\Sigma_u^-$ \\ 
		$4s$ &  4512   & $0^{-+}$ & $1^{--}$ & $\Sigma_g^+$ \\ 
		$2p_0$   & 4920 & $0^{++}$ & $1^{+-}$ & $\Sigma_u^-$ \\ 
		$3p_0$  & 5299 & $0^{++}$ & $1^{+-}$ & $\Sigma_u^-$ \\ 
		$4p_0$    & 5642 & $0^{++}$ & $1^{+-}$ & $\Sigma_u^-$ \\ \hline
		$1p$ &  3494   & $1^{+-}$ & $(0,1,2)^{++}$ & $\Sigma_g^+$ \\ 
		$2p$ &  3968   & $1^{+-}$ & $(0,1,2)^{++}$ & $\Sigma_g^+$ \\ 
		$1(s/d)_1$   & 4011 & $1^{--}$ & $(0,1,2)^{-+}$ & $\Pi_u\Sigma_u^-$ \\ 
		$1p_1$   & 4145 & $1^{++}$ & $(0,1,2)^{+-}$ & $\Pi_u$ \\ 
		$2(s/d)_1$   & 4355 & $1^{--}$ & $(0,1,2)^{-+}$ & $\Pi_u\Sigma_u^-$ \\ 
		$3p$ &  4369   & $1^{+-}$ & $(0,1,2)^{++}$ & $\Sigma_g^+$ \\
		$2p_1$ &    4511 & $1^{++}$ & $(0,1,2)^{+-}$ & $\Pi_u$ \\ 
		$3(s/d)_1$ &   4692 & $1^{--}$ & $(0,1,2)^{-+}$ & $\Pi_u\Sigma_u^-$ \\ 
		$4(s/d)_1$ &   4718 & $1^{--}$ & $(0,1,2)^{-+}$ & $\Pi_u\Sigma_u^-$ \\ 
		$4p$ &  4727 &   $1^{+-}$ & $(0,1,2)^{++}$ & $\Sigma_g^+$ \\ 
		$3p_1$&    4863 & $1^{++}$ & $(0,1,2)^{+-}$ & $\Pi_u$ \\ 
		$5(s/d)_1$ &   5043 & $1^{--}$ & $(0,1,2)^{-+}$ & $\Pi_u\Sigma_u^-$ \\ 
		$5p$ & 5055 &   $1^{+-}$ & $(0,1,2)^{++}$ & $\Sigma_g^+$ \\ \hline
		$1d$ & 3793 &   $2^{-+}$ & $(1,2,3)^{--}$ & $\Sigma_g^+$ \\ 
		$2d$ & 4210 &   $2^{-+}$ & $(1,2,3)^{--}$ & $\Sigma_g^+$ \\ 
		$1(p/f)_2$ &  4231 & $2^{++}$ & $(1,2,3)^{+-}$ & $\Pi_u\Sigma_u^-$ \\ 
		$1d_2$  &   4334 & $2^{--}$ & $(1,2,3)^{-+}$ & $\Pi_u$ \\ 
		$2(p/f)_2$  &   4563 & $2^{++}$ & $(1,2,3)^{+-}$ & $\Pi_u\Sigma_u^-$ \\ 
		$3d$  & 4579 &   $2^{-+}$ & $(1,2,3)^{--}$ & $\Sigma_g^+$ \\ 
		$2d_2$ &   4693 & $2^{--}$ & $(1,2,3)^{-+}$ & $\Pi_u$ \\ 
		$3(p/f)_2$ &   4886 & $2^{++}$ & $(1,2,3)^{+-}$ & $\Pi_u\Sigma_u^-$ \\ 
		$4d$  & 4916 &  $2^{-+}$ & $(1,2,3)^{--}$ & $\Sigma_g^+$ \\ 
		$4(p/f)_2$  &   4923 & $2^{++}$ & $(1,2,3)^{+-}$ & $\Pi_u\Sigma_u^-$ \\ 
		$3d_2$ &   5036 & $2^{--}$ & $(1,2,3)^{-+}$ & $\Pi_u$ \\ \hline
	\end{tabular}
	\caption{ Charmonium ($\Sigma_g^+$) and hybrid charmonium ($\Pi_u\,,\Sigma_u^-$) energy spectrum computed with 
	$m_c=1.47$ GeV. Masses are in MeV. States which only differ by the heavy quark spin $(S=0,1)$ are degenerated.
	$N$ is the principal quantum number, $L$ the orbital angular momentum of the heavy quarks, $J$ is $L$ plus the total angular momentum of the gluons, $S$ the spin of the heavy quarks and ${\cal J}$ is the total angular momentum. 
	For quarkonium, $J$ coincides with $L$ and it is not displayed.
	The last column shows the relevant potentials for each state. %The $(s/d)_1$, $p_1$, $p_0$, $(p/f)_2$ and $d_2$ states are named $H_1$, $H_2$, $H_3$, $H_4$ and $H_5$ respectively in \cite{Berwein:2015vca}. 
	}
	\label{cEspectrum}
\end{table}

	\begin{table}[htbp]
		\centering
		\begin{tabular}{|c|c|c|c|c|}
			\hline 
			&                        &  $S=0$ & $S=1$ & \\
			$NL_J$  & $M$%_{b\bar{b}}$ & $M_{b\bar{b}g}$ 
			& $\mathcal{J}^{PC}$ & $\mathcal{J}^{PC}$ & $\Lambda^{\epsilon}_{\eta}$ \\ \hline \hline
			$1s$ & 9442 &   $0^{-+}$ & $1^{--}$ & $\Sigma_g^+$ \\ 
			$2s$ & 10009 &   $0^{-+}$ & $1^{--}$ & $\Sigma_g^+$ \\ 
			$3s$ & 10356 &   $0^{-+}$ & $1^{--}$ & $\Sigma_g^+$ \\ 
			$4s$ & 10638 &   $0^{-+}$ & $1^{--}$ & $\Sigma_g^+$ \\ 
			$1p_0$  &   11011 & $0^{++}$ & $1^{+-}$ & $\Sigma_u^-$ \\ 
			$2p_0$  &   11299 & $0^{++}$ & $1^{+-}$ & $\Sigma_u^-$ \\ 
			$3p_0$  &   11551 & $0^{++}$ & $1^{+-}$ & $\Sigma_u^-$ \\ 
			$4p_0$  &   11779 & $0^{++}$ & $1^{+-}$ & $\Sigma_u^-$ \\ \hline
			$1p$ & 9908 &   $1^{+-}$ & $(0,1,2)^{++}$ & $\Sigma_g^+$ \\ 
			$2p$  & 10265 &   $1^{+-}$ & $(0,1,2)^{++}$ & $\Sigma_g^+$ \\ 
			$3p$ & 10553 &   $1^{+-}$ & $(0,1,2)^{++}$ & $\Sigma_g^+$ \\ 
			$1(s/d)_1$  &  10690 & $1^{--}$ & $(0,1,2)^{-+}$ & $\Pi_u\Sigma_u^-$ \\ 
			$1p_1$  &  10761 & $1^{++}$ & $(0,1,2)^{+-}$ & $\Pi_u$ \\ 
			$4p$ & 10806 &   $1^{+-}$ & $(0,1,2)^{++}$ & $\Sigma_g^+$ \\ 
			$2(s/d)_1$  &   10885 & $1^{--}$ & $(0,1,2)^{-+}$ & $\Pi_u\Sigma_u^-$ \\ 
			$2p_1$  &   10970 & $1^{++}$ & $(0,1,2)^{+-}$ & $\Pi_u$ \\ 
			$5p$  & 11035 &   $1^{+-}$ & $(0,1,2)^{++}$ & $\Sigma_g^+$ \\ 
			$3(s/d)_1$  &   11084 & $1^{--}$ & $(0,1,2)^{-+}$ & $\Pi_u\Sigma_u^-$ \\ 
			$4(s/d)_1$  &   11156 & $1^{--}$ & $(0,1,2)^{-+}$ & $\Pi_u\Sigma_u^-$ \\ 
			$3p_1$ &   11175 & $1^{++}$ & $(0,1,2)^{+-}$ & $\Pi_u$ \\
			$6p$  & 11247   & $1^{+-}$ & $(0,1,2)^{++}$ & $\Sigma_g^+$ \\ 
			$5(s/d)_1$ &   11284 & $1^{--}$ & $(0,1,2)^{-+}$ & $\Pi_u\Sigma_u^-$ \\ \hline
			$1d$ & 10155  & $2^{-+}$ & $(1,2,3)^{--}$ & $\Sigma_g^+$ \\ 
			$2d$ & 10454   & $2^{-+}$ & $(1,2,3)^{--}$ & $\Sigma_g^+$ \\ 
			$3d$ & 10712   & $2^{-+}$ & $(1,2,3)^{--}$ & $\Sigma_g^+$ \\
			$1(p/f)_2$   & 10819 & $2^{++}$ & $(1,2,3)^{+-}$ & $\Pi_u\Sigma_u^-$ \\ 
			$1d_2$  &   10870 & $2^{--}$ & $(1,2,3)^{-+}$ & $\Pi_u$ \\ 
			$4d$ & 10947  & $2^{-+}$ & $(1,2,3)^{--}$ & $\Sigma_g^+$ \\ 
			$2(p/f)_2$  &   11005 & $2^{++}$ & $(1,2,3)^{+-}$ & $\Pi_u\Sigma_u^-$ \\ 
			$2d_2$ &   11074 & $2^{--}$ & $(1,2,3)^{-+}$ & $\Pi_u$ \\ 
			$5d$  & 11163   & $2^{-+}$ & $(1,2,3)^{--}$ & $\Sigma_g^+$ \\ 
			$3(p/f)_2$   & 11197 & $2^{++}$ & $(1,2,3)^{+-}$ & $\Pi_u\Sigma_u^-$ \\ 
			$3d_2$ &   11275 & $2^{--}$ & $(1,2,3)^{-+}$ & $\Pi_u$ \\ 
			$4(p/f)_2$ &  11291 & $2^{++}$ & $(1,2,3)^{+-}$ & $\Pi_u\Sigma_u^-$ \\ \hline
		\end{tabular}
		\caption{ Bottomonium  ($\Sigma_g^+$) and hybrid bottomonium ($\Pi_u\,,\Sigma_u^-$) energy spectrum computed with $m_b=4.88$ GeV. 
		Masses are in MeV. States which only differ by the heavy quark spin $(S=0,1)$ are degenerated.
	$N$ is the principal quantum number, $L$ the orbital angular momentum of the heavy quarks, $J$ is $L$ plus the total angular momentum of the gluons, $S$ the spin of the heavy quarks and ${\cal J}$ is the total angular momentum. 
	For quarkonium, $J$ coincides with $L$ and it is not displayed.
	The last column shows the relevant potentials for each state. %The $(s/d)_1$, $p_1$, $p_0$, $(p/f)_2$ and $d_2$ states are named $H_1$, $H_2$, $H_3$, $H_4$ and $H_5$ respectively in \cite{Berwein:2015vca}. 
	}
	\label{bEspectrum}
	\end{table}

The lightest hybrid multiplet corresponds to the $1\,(s/d)_1$ quantum number ($4011$ MeV for charmonium and $10690$ MeV for bottomonium). The remaining hierarchy from lighter to heavier reads 
$1\, p_1$, $1\, (p/f)_2$, $1\, d_2$, $2\,(s/d)_1$, $\dots$ 
The $XYZ$ states that fit in our hybrid spectrum are displayed in Table $3$. %\ref{sxyz}. 
These numbers are accurate up to ${\cal O}(\lQ^2/m_Q)$ corrections, namely about $110$ MeV for charmonium and $33$ MeV for bottomonium.
{%\scriptsize 
\tiny
\begin{table}[htbp]
\label{sxyz}
%\centering
	\begin{tabular}{|c|c|c|c|c|}
		\hline
		State & M & $J^{PC}$ & XYZ  %& M$_{exp}$ & $\Gamma_{exp}$ 
		& $J^{PC}_{exp}$ \\ \hline
	%	1s1 & (n=1)4486 & $0^{++}$,$1^{+-}$ & Z(4430) & $4433\pm5$ & $45^{+35}_{-18}$ & $?^{??}$ \\ \hline
		$1(s/d)_1$ & %(n=1/1)
		4011 & $1^{--}$,$(0,1,2)^{-+}$ & Y(4008) %& $4008^{+121}_{-49}$ & $226\pm97$ 
		& $1^{--}$ \\ \hline
		$1p_1$ & %(n=1)
		4145 & $1^{++}$,$(0,1,2)^{+-}$ & Y(4140) %& $4144,5\pm2,6$ & $15^{+11}_{-7}$ 
		& $1^{++}$ \\ 
		\multicolumn{1}{|l|}{} & \multicolumn{1}{l|}{} & \multicolumn{1}{l|}{} & X(4160) %& $4156^{+29}_{-25}$ & $139^{+113}_{-65}$ 
		& $?^{??}$\\ \hline
		\multicolumn{1}{|l|}{} & \multicolumn{1}{l|}{} & \multicolumn{1}{l|}{} & X(4320) %& $4320{\pm 17}$ & $101{\pm 30}$ 
		& $1^{--}$ \\ 
		\multicolumn{1}{|l|}{} & \multicolumn{1}{l|}{} & \multicolumn{1}{l|}{} & X(4350) %& $4351{\pm 5}$ & $13^{+18}_{-10}$ 
		& $?^{?+}$\\
		$2(s/d)_1$ & %(n=2/2)
		4355 & $1^{--}$,$(0,1,2)^{-+}$ & Y(4360) %& $4361\pm13$ & $74\pm18$ 
		& $1^{--}$ \\
		\multicolumn{1}{|l|}{} & \multicolumn{1}{l|}{} & \multicolumn{1}{l|}{} & Y(4390) %& $4391{\pm 6}$ & $139{\pm 16}$ 
		& $1^{--}$\\ \hline
		$1p_0$ & %(n=1)
		4486 & $0^{++}$,$1^{+-}$ & X(4500) %& $4506^{+16}_{-19}$ & $92^{+30}_{-29}$ 
		& $0^{++}$ \\ \hline
		$3(s/d)_1$ & %(n=2/3)
		4692 & $1^{--}$,$(0,1,2)^{-+}$ & Y(4660) %& $4664\pm12$ & $48\pm15$ 
		& $1^{--}$ \\ 
		\multicolumn{1}{|l|}{} & \multicolumn{1}{l|}{} & \multicolumn{1}{l|}{} & X(4630) %& $4634^{+9}_{-11}$ & $92^{+41}_{-32}$ 
		& $1^{--}$ \\% \hline
		%$1(p/f)_2$ & %(n=1/1)
		%4231 & $2^{++}$,$(1,2,3)^{+-}$ & Y(4274) & $4274,4^{+8,4}_{-6,7}$ & $32^{+22}_{-15}$ & $?^{?+}$ \\
		\hline \hline
		$2(s/d)_1$ & %(n=2/2)
		10885 & $1^{--}$,$(0,1,2)^{-+}$ & $Y_b$(10890) %& $10888,4\pm3$ & $30,7^{+8,9}_{-7,7}$ 
		& $1^{--}$ \\ \hline
	\end{tabular}
	\hfill
	\caption{Hybrid states in our spectrum with masses and quantum numbers compatible with charmonium (above) and bottomonium (below) $XYZ$ resonances. }
	\end{table}
	}
	%\begin{itemize}
%\item 
C-parity implies that only spin zero hybrids would have been observed, except for $X(4350)$.
%\item 
However, decays to spin one quarkonium states have been observed for all spin zero $1^{--}$ states above, except for $X(4630)$ and $Y(4390)$, which disfavors the hybrid interpretation due to spin symmetry. We show in the following section how this problem can be overcome.

\subsection{Mixing}

Mixing with quarkonium is in principle an $1/m_Q$ suppressed effect. However, if there is a quarkonium state with mass close to a hybrid one's, it may become a leading order effect. The symmetries of the quarkonium (\ref{symS}) and the hybrid fields (\ref{symH}) imply that the mixing at order $1/m_Q$ is controlled by a single term,
\be
\label{mix}
{\cal L}_{\rm mixing} = {\rm tr}\left[%S^\dagger L^{i} V_L^{ij} H^j
%\right]
  S^\dagger  V_S^{ij} \left\{ \sigma^i \, , H^j\right\} 
%\right
+ {\rm h.c.}\right]\,.
\ee
Indeed, the quarkonium field $S=S({\bf R},{\bf r}, t)$ transforms like {\bf H} under heavy quark spin symmetry 
and as follows under the discrete symmetries \cite{Brambilla:2004jw},
\bea
\label{symS}
P: & \, S({\bf R},{\bf r}, t)& \rightarrow -S(-{\bf R},-{\bf r}, t)\nn\\
T: &\, S({\bf R},{\bf r}, t)& \rightarrow \sigma^2 S({\bf R},{\bf r}, -t)\sigma^2\\
C: &\,  S({\bf R},{\bf r}, t)& \rightarrow \sigma^2S^T({\bf R},-{\bf r}, t)\sigma^2  \,.\nn
\eea
The term (\ref{mix}) mixes spin zero (one) hybrids with spin one (zero) quarkonium, and may be the source of large spin symmetry violations. $V_S^{ij}$ is unknown at the moment, but it can be easily evaluated on the lattice. Explicit formulas can be found in ref. \cite{Oncala:2017hop}. We can work out, however, its short distance behavior from the weak coupling regime of pNRQCD \cite{Brambilla:2002nu}, and its long distance one from the QCD EST \cite{Luscher:2004ib}. We have, in general,
\be
V_S^{ij}=(\delta^{ij}-{\hat r}^i{\hat r}^j) V_S^{\Pi} + {\hat r}^i{\hat r}^jV_S^{\Sigma}\,.
\ee
$V_S^{\Pi}$ and $V_S^{\Sigma}$ are proportional to the NRQCD matching coefficient of the chromomagnetic interaction $c_F$, which is known at three loops \cite{Grozin:2007fh}. In the following, we approximate it by its tree level value $c_F=1$. Then, we obtain for the short distance behavior,
\be
V_S^{\Pi}(r) \sim V_S^{\Sigma}(r) \rightarrow \pm\frac{\lambda^2}{m_Q}\,,
\ee
where $\lambda$ is a constant ${\cal O}(\lQ)$, and for the long distance one,
\be
V_S^{\Sigma}(r)\rightarrow %\frac{16}{\sqrt{2}}
-\frac{\pi^2 g\Lambda'''}{m_Q\kappa r^3}\, ,\; V_S^{\Pi}(r) \rightarrow
\sqrt{\frac{\pi^3}{\kappa}}\frac{g\Lambda'}{2m_Q r^2}\,.
\ee
$\kappa=\kappa_g$ is the string tension given in (\ref{st}), and $\Lambda'$ and $\Lambda'''$ are constants ${\cal O}(\lQ)$ that also appear in the long distance behavior of the quarkonium spin-orbit and tensor potentials \cite{PerezNadal:2008vm,Brambilla:2014eaa}. We obtain from fits to the lattice data for those potentials in ref. \cite{Koma:2009ws},
\be
%\kappa\sim 0.187 {\rm GeV^2}\, ,\, 
g\Lambda'\sim -59 \,{\rm MeV} \, ,\, g\Lambda'''\sim \pm 230\, {\rm MeV}\,.
\ee
We model the mixing potential with simple interpolations that reproduce the correct short and long distance limits, and allow for a sign flip between the two limits,
\bea
\label{inter}
 &&V_S^\Pi[\pm -](r)=\frac{\lambda^2}{m_Q}\left(\frac{\pm 1-(\frac{r}{r_\Pi})^2}{1+(\frac{r}{r_\Pi})^4}\right)\,\nn\\&&
 V_S^\Sigma[\pm\pm](r)=\frac{\lambda^2}{m_Q}\left(\frac{\pm 1\pm (\frac{r}{r_\Sigma})^2}{1+(\frac{r}{r_\Sigma})^5}\right)\\
&&\, r_\Pi=(\frac{\vert g\Lambda'\vert\pi^{\frac{3}{2}}}{2\lambda^2\kappa^{\frac{1}{2}}})^{\frac{1}{2}} \,,\, r_\Sigma=(\frac{\vert g\Lambda'''\vert\pi^2}{\lambda^2\kappa})^{\frac{1}{3}}\,.\nn
\eea

When (\ref{mix}) is written in the $\vert SLJ\j\m \rangle$ basis (for quarkonium $L=J$ always) we obtain two sets of equations. The firts (second) set mixes $S=0$ ($S=1$) hybrids with $S=1$ ($S=0$) and spans a 6 (10) dimensional subspace. 
We write $S_{S\, \j\m}^L$ for the quarkonium wave functions  (in the case of $S=0$ the upper index is omitted since $L=\j$ always) and $P_{S\, \j\m}^{LJ}$ for the hybrid ones (in the case $S=0$ the index $J$ is omitted 
since $J=\j$ always). We use $0,\pm$ 
as a shorthand for $J=\j,\j\pm 1$ and for $L=J,J\pm 1$. The quantum mechanical Hamiltonian of the first set has 
a two dimensional invariant subspace spanned by $(S_{1\, \j\m}^0, P_{0\, \j\m}^0)$ and a four dimensional one spanned by $(S_{1\, \j\m}^+, S_{1\, \j\m}^-, P_{0\, \j\m}^+, P_{0\, \j\m}^-)$. The quantum mechanical Hamiltonian 
of the second set has a six dimensional invariant subspace spanned by $(S_{0\, \j\m}, P_{1\, \j\m}^{++}, P_{1\, \j\m}^{-+}, P_{1\, \j\m}^{+-}, P_{1\, \j\m}^{--}, P_{1\, \j\m}^{00})$, and a trivial four dimensional one spanned 
by $(P_{1\, \j\m}^{-0}, P_{1\, \j\m}^{+0}, P_{1\, \j\m}^{0-}, P_{1\, \j\m}^{0+})$, which corresponds to hybrids with exotic $J^{PC}$ that do not mix with quarkonium. The specific form of these Hamiltonians is given in formulas 
(51)-(54) of ref. \cite{Oncala:2017hop}.

We focus first on the hybrid $S=0$ sector of charmonium. We scan $\lambda= 100, 300, 600$ MeV for all possible sign combinations in (\ref{inter}) and solve the eigenvalue equations for each case. We observed from the results that
$V_S^\Pi [+-]$ together with $V_S^\Sigma [++]$ and $\lambda=600$ MeV produce the maximum mixing (see Table IV of ref. \cite{Oncala:2017hop}). With this choice, $Y(4008)$, $Y(4360)$  and $Y(4660)$ contain $29\%$, $35\%$ and $17\%$ of spin one quarkonium. Hence, 
the spin symmetry violating decays of $Y(4008)$, $Y(4360)$ and $Y(4660)$ are qualitatively explained. We keep this choice of signs and $\lambda$ fix for the remaining calculations both in the charmonium and bottomonium cases. 
The complete results are presented in Tables IV and VI-XII of ref. \cite{Oncala:2017hop}.  We only display here, in Table $4$, %\ref{xyz}, 
the $XYZ$ states that can be identified in our spectrum. The mixing produces small shifts 
in the spectrum, of a few MeV, and also induces small hyperfine splittings. However, the mixing between hybrids and quarkonia is large in a number of cases, some of them of phenomenological relevance, as we will discuss 
in Sec. \ref{discussion}.  Needless to say that it would be important that lattice calculations confirm the signs and size of the mixing potentials above. The calculation remains accurate up to ${\cal O}(\lQ^2/m_Q)$ corrections.

\begin{table}[htbp]
%\centering
	\begin{tabular}{|c|c|c|c|}
		\hline
	Resonance & $J^{PC}_{exp}$ & Assignement & Mass (MeV) %& Observations 
	\\ \hline
	X(3823) & $2^{--}$ & $1d$ & 3792  \\
	X(3872) & $1^{++}$ & $2p$ & 3967  \\
	X(3915) & ${0 \,{\rm or}\, 2} ^{++}$ & $2p$ & 3968  \\
  X(3940) & $?^{??}$ & $2p$ & 3968  \\
	Y(4008) & $1^{--}$ & $1{s/d}_1$ & 4004 %& mixing 
	\\
	X(4140) & $1^{++}$ & ?? & ?? %& 	$1p_1$ does not decay to quarkonium
	\\
	X(4160) & $?^{??}$ & $1p_1$ & 4146  \\
	Y(4220) & $1^{--}$ & $2d$ & 4180 %&  $Y(4260)\to Y(4220)$, mixing 
	\\
	X(4230) & $1^{--}$ & $2d$ & 4180 %&  $X(4230)=Y(4220)$, mixing
	\\
	Y(4274) & $1^{++}$ & $3p$ & 4368 %&  $X(4230)=Y(4220)$, mixing
	\\
X(4350) & $?^{?+}$ & $2(s/d)_1$ or 3p & 4355 or 4369  \\
	Y(4320) & $1^{--}$ & $2(s/d)_1$ & 4366  %& mixing 
	\\
	Y(4360) & $1^{--}$ & $2(s/d)_1$ & 4366  %& $Y(4360)=Y(4320) ?$
	\\
	X(4390) & $1^{--}$ & $2(s/d)_1$ & 4366  %& $ Y(4390)=Y(4360) ?$
	\\
	X(4500) & $0^{++}$ & $1p_0$ & 4566 %& not enough mixing 
	\\
	Y(4630) & $1^{--}$ & $3d$ & 4559  \\
Y(4660) & $1^{--}$ & $3(s/d)_1$ & 4711 %&  mixing 
\\
	X(4700) & $0^{++}$ & $4p$ & 4703 \\ \hline
	%$\Upsilon$(10860) & $1^{--}$ & $5s$ & 10881% &  mixing
	%\\
	Y$_{\rm b}$(10890) & $1^{--}$ & $2(s/d)_1$ & 10890 %&  mixing
	%\\
	%$\Upsilon$(11020) & $1^{--}$ & $4d$ & 10942
	\\ \hline
	\end{tabular}
	\label{xyz}
	\caption{The identification of isospin zero $XYZ$ charmonium (above) and bottomonium (below) resonances with states in our spetrum. The last column shows the masses obtained in our calculation.}
	\end{table}

\section{Decay}

The energy gap of some hybrid states to the lower lying quarkonium states is around $1$ GeV or bigger, which suggests that these decays can be estimated perturbatively, if the accompanying light mesons are left unspecified, namely as semi-inclusive decays. In the effective theory language, the lower lying quarkonia can be integrated out producing and imaginary
contribution to the Lagragian that eventually leads to the decay width. Using weak coupling pNRQCD at leading non-trivial order in the multipole expansion we obtain that hybrid states with $L=J$ do not decay to quarkonium. This selection rule eliminates $X(4140)$ as a candidate of the charmonium $1\,p_1$ hybrid state. For $L\not= J$ we can give reliable estimates for a small number of decays for charmonium and a somewhat larger number for bottomonium, which are displayed in Table \ref{decayW}.%III of ref. \cite{Oncala:2017hop}.

\begin{table}[htbp]
	\centering
	\begin{tabular}{|c|c|c|c|c|c|}
		\hline

			$NL_J \rightarrow N'L'$ & $\Delta E$ (MeV)
%& $\langle r\rangle 
%_{mn}$ 
 %& ${\vert\Delta E \langle r\rangle 
%_{mn}\vert}$ & $\als (\Delta E)$ 
& {$\Gamma$ (MeV) }  \\ \hline
					
		$1p_0\rightarrow 2s$ & {808} %& {0.40} & {0.32} & 0.41 
		& {7.5(7.4)} \\ 
		$2(s/d)_1\rightarrow 1p$ & 861 %& 0.63 & 0.54 & 0.39 
		& 22(19) \\ 
                $4(s/d)_1\rightarrow 1p$ & 1224 %& 0.42 & 0.51 & 0.33 
								& 23(15) \\ 
		\hline
		\hline
		$1p_0\rightarrow 1s$ & {1569} %& {-0.42} & {0.65} & 0.29
		& {44(23)}   \\ 
		$1p_0\rightarrow 2s$ & {1002} %& {0.43} & {0.43} & 0.36 
		& {15(9)}   \\  
		$2p_0\rightarrow 2s$ & {1290} %& {-0.14} & {0.18} & 0,32 
		&{2.9(1.3)}   \\ 
		$2p_0\rightarrow 3s$ & {943} %& {0.46} & {0.44} & 0.37 
		& {15(12)}   \\ 
                $4p_0\rightarrow 1s$ & {2337} %& {0.27} & {0.63} & 0.25 
								& {53(25)}   \\
                 $4p_0\rightarrow 2s$ & {1770} %& {0.23} & {0.40} & 0.28 
								& {18(7)}   \\
                $4p_0\rightarrow 3s$ & {1423} %& {0.19} & {0.28} & 0.31 
								& {7.4(4.1)}   \\
		{$2(s/d)_1\rightarrow 1p$} & {977} %& {0.47} & {0.46} & 0.37 
		& {17(8)}  \\ 
                 {$3(s/d)_1\rightarrow 1p$} & {1176} %& {0.49} & {0.58} & 0.33 
								& {29(14)}  \\ 
                 {$3(s/d)_1\rightarrow 2p$} & {818} %& {0.32} & {0.26} & 0.40 
								& {5(3)}  \\ 
                 {$4(s/d)_1\rightarrow 2p$} & {891} %& {-0.74} & {0.66} & 0.39 
								& {33(25)}  \\
                 {$5(s/d)_1\rightarrow 1p$} & {1376} %& {-0.31} & {0.43} & 0.31 
								& {18(7)}  \\ 
                 {$5(s/d)_1\rightarrow 2p$} & {1018} %& {-0.41} & {0.42} & 0.36 
								& {14(8)}  \\
		\hline
	\end{tabular}
	\caption{Decay widths for hybrid charmonium (above) and bottomonium (below) to lower lying charmonia and bottomonia respectively. %$m=NL_J$, $n=N'L'$, $\Delta E\equiv \Delta E_{mn}$ and $\Gamma$ are in MeV, and 
%$\langle r\rangle _
%{mn}$ in GeV$^{-1}$. $\als (\Delta E)$ is the one-loop running coupling constant at the scale 
$\Delta E$ is the energy difference between the hybrid and quarkonium states. Mixing with quarkonia has been neglected. %We only display results
	%for which $\Delta E > 800$MeV and ${\vert\Delta E \langle r\rangle _{mn}\vert} < 0.7$. 
	The error (in brackets) is discussed in ref. \cite{Oncala:2017hop}. %includes higher orders in $\als$ and in the multipole expansion, as well as  
%the average of the linear term in the Cornell potential in order to account for the difference between weak and strong coupling regimes.
}
	\label{decayW}
\end{table}

\section{Hyperfine splittings}

Hyperfine splittings appear at ${\cal O}(1/m_Q)$  in hybrids rather than at ${\cal O}(1/m_Q^2)$ as in quarkonium. Furthermore, at leading order, they are controlled by a single term in the Lagrangian {\cite{Oncala:2017hop},
\bea
\label{hfs}
&&i\epsilon^{ijk}V^S(r)\, {\rm tr}\left(H^{i\dagger}\left[\sigma^k , H^j\right]\right) 
\eea 
This allows to put forward results that are independent of the form of $V^S(r)$ \cite{Pere}. Let us specify first the quantum mechanical Hamiltonian, $H_{hf}$, following from (\ref{hfs}) in the $\vert SLJ\j\m\rangle$ basis. %We write $0,\pm$ 
%as a shorthand for $J=\j,\j\pm 1$ and for $L=J,J\pm 1$ and label the radial wave functions as $P_{S\, \j\m}^{LJ}$. 
In this basis, $H_{hf}$ contains diagonal and off-diagonal terms, and it is non-vanishing on $S=1$ states only. It is also proportional to $-2V^S(r)$, a global factor that we will skip in following. $H_{hf}$ is a nine by nine matrix
that decomposes into two one dimensional boxes, two two dimensional boxes and one three dimensional box. The two one dimensional boxes correspond to $P_{1\, \j\m}^{++}$ and $P_{1\, \j\m}^{--}$ respectively and have eigenvalue $1$.
 The two two dimensional boxes correspond to the subspaces spanned by $(P_{1\, \j\m}^{+0}\,,P_{1\, \j\m}^{0+})$ and $(P_{1\, \j\m}^{0-}\,, P_{1\, \j\m}^{-0})$, they read
\be
\begin{pmatrix}
 		\!\frac{1}{\j+1}  & \frac{\sqrt{(\j+2)\j}}{\j+1}                    \\
 		\frac{\sqrt{(\j+2)\j}}{\j+1} & 	-\frac{1}{\j+1}                    \\
 	\end{pmatrix}
	\,\,,\,\,
	\begin{pmatrix}
 		\!\frac{1}{\j}  & \frac{\sqrt{\j^2-1}}{\j}                    \\
 		\frac{\sqrt{\j^2-1}}{\j} & 	-\frac{1}{\j}                    \\
 	\end{pmatrix}
	\ee
	The three dimensional box corresponds to the subspace spanned by $(P_{1\, \j\m}^{+-}\,,P_{1\, \j\m}^{00}\,,P_{1\, \j\m}^{-+})$, it reads,
	\be
	\begin{pmatrix}
 		\!-\frac{\j-1}{\j}  & \frac{\j+1}{\j}\frac{\sqrt{2\j-1}}{\sqrt{2\j+1}} & 0                    \\
 		\frac{\j+1}{\j}\frac{\sqrt{2\j-1}}{\sqrt{2\j+1}} & 	-\frac{1}{\j (\j+1)}  & \frac{\j}{\j+1}\frac{\sqrt{2\j+3}}{\sqrt{2\j+1}}                  \\
		0 & \frac{\j}{\j+1}\frac{\sqrt{2\j+3}}{\sqrt{2\j+1}} & -\frac{\j+2}{\j+1}\\
 	\end{pmatrix}
	\ee
At first order in perturbation theory, only the diagonal terms contribute. They lead to the following mass formulas for a quadruplet $J$
\be
 \frac{M_{1\, J+1}-M_{0\, J}}{M_{1\, J}-M_{0\, J}}=-J \,,\quad \frac{M_{1\, J-1}-M_{0\, J}}{M_{1\, J}-M_{0\, J}}=J+1\,,
\ee
where $M_{S\, \j}$ denotes the mass of a given state in the quadruplet ($\j=J$ for $S=0$, and $\j=J\,, J\pm 1$ for $S=1$). For the lower lying quadruplets, we have,
\bea
\label{massf}
(s/d)_1:& M_{2^{-+}}+ M_{0^{-+}} &= M_{1^{-+}}+M_{1^{--}}\nn\\
p_1:& M_{2^{+-}}+ M_{0^{+-}} &= M_{1^{+-}}+M_{1^{++}}\nn\\
(p/f)_2:& M_{3^{+-}}+ M_{1^{+-}} &= M_{2^{+-}}+M_{2^{++}}\\
d_2:& M_{3^{+-}}+ M_{1^{+-}} &= M_{2^{+-}}+M_{2^{--}}\nn
\eea
The three first formulas can be checked against recent lattice data from HSC collaboration \cite{Cheung:2016bym}. The difference between the lhs and rhs is $14$ MeV, $45$ MeV and $39$ MeV
respectively. These figures fall within the expected size of the ${\cal O}(\lQ^3/m_Q^2)$ corrections, and hence (\ref{massf}) is consistent with the data of ref. \cite{Cheung:2016bym}.
Note, however, that 
the off-diagonal terms in $H_{hf}$ mix different $J$ multiplets, and may lead to enhance ${\cal O}(\lQ^3/m_Q^2)$ contributions to the hyperfine splittings if there are multiplets with similar masses, in an analogous way as the mixing with quarkonia leads to enhanced spin symmetry violations.

\section{Discussion}
\label{discussion}

All the isospin zero $XYZ$ resonances fit well in our spectrum, either as a quarkonium or hybrid states, except for $X(4140)$. This resonance could in principle be assigned to the $1p_1$ spin zero charmonium hybrid. However the selection rule we find that $L=J$ hybrids do not decay to quarkonium at leading order, disfavors this assignment. We end up assigning
 the $X(4160)$ to the $1p_1$ spin zero hybrid, and since there is no other state with $1^{++}$ quantum numbers nearby in our spectrum, $X(4140)$ is left with no assignment.

The mixing of spin zero hybrids with spin one quarkonia is important for the assignments of $Y(4008)$, $Y(4220)$, $X(4230)$, $Y(4320)$, $Y(4360)$, $Y(4660)$ and $Y_b (10890)$ as hybrid states, since otherwise spin symmetry violating decays of these resonances would have been observed. It is also important in order to understand why the spin symmetry violating transitions of $\Upsilon (10860)$ to $P$-wave states are of the same order as the ones that respect spin symmetry.

In the charmonium spectrum there are too many $1^{--}$ resonances. The wide $Y(4008)$ resonance was observed by Belle \cite{Yuan:2007sj,Liu:2013dau}, but it has not been confirmed by Babar \cite{Lees:2012cn} or BESIII \cite{Ablikim:2016qzw}. If we assume that this resonance is not there, then there is no assignment for $1(s/d)_1$ spin zero state, unless we assign it to the $\psi (4040)$. The fact that the $1(s/d)_1$ spin zero state has about a $30\%$ mixing with spin one quarkonium may justify the assignment. Then the $3s$ state would be the $\psi (4160)$ and the $2d$ state the $X(4230)/Y(4220)$. The $X(4230)$ and the $Y(4220)$ resonances have compatible parameters and must be identified for them both to be compatible with our spectrum. For the spin zero $2(s/d)_1$ state, there are three competing resonances that should be identified, $Y(4320)$, $Y(4360)$, and $Y(4390)$. Indeed the decay widths and masses are compatible within $1\sigma$, except for the mass of $Y(4390)$. If this picture is correct, below the $Y(4660)$ resonance there would only be the $3d$ state to be discovered, around $4560$ MeV. Furthermore, from Table \ref{decayW} and the mixing pattern, we are able to estimate the following decay widths,
\be
\Gamma \left(Y(4320/4360/4390) \to h_c + {\rm l.h.}\right)= 14 (12)\, {\rm MeV} \,
\ee  
\be
\Gamma \left(X(4230)/Y(4220) \to h_c + {\rm l. h.}\right)= 17 (15)\, {\rm MeV} \,, \nn
\ee 
where l.h. stands for light hadrons. %Analogously, for the $X(4230)/Y(4220)$ state we have,

Concerning $1^{--}$ bottomonium resonances, all of them fit in our spectrum. In addition there should be three states to be discovered below the $\Upsilon (10860)$, the $2d$, the $1(s/d)_1$ and the $3d$, around $10440$ MeV, $10690$ MeV and $10710$ MeV respectively. Analogously to the charmonium case above, we can also put forward some estimates for the decay widths,
\bea
\Gamma \left(\Upsilon(10860) \to h_b + {\rm l.h.}\right) &=& 3 (1)\, {\rm MeV} \nn\\
\Gamma \left(Y_b(10890) \to h_b + {\rm l.h.}\right) &=& 13 (6)\, {\rm MeV}\, .
\eea

Let us finaly discuss the important question on how the lattice potentials we use (Fig. \ref{kuti})
may change in the case $n_f=3$ (three light quarks). 
We know that $\Sigma_g^+$ does not change
much and this is also so for $\Pi_u$ \cite{Bali:2000vr}, at least up to moderately large distances.
 We do not have such an information about $\Sigma_u^-$, but there is no reason 
to expect a different behavior. Two major qualitative features arise though. The first one is 
the appearance 
of heavy-light meson pairs, which amount to roughly horizontal lines at the threshold energies in Fig. \ref{kuti}. 
These states interact with the remaining potentials already at leading order, and may in principle 
produce important distortions with respect to the $n_f=0$ case.
However, we know how they cross talk to the $\Sigma_g^+$ potential \cite{Bali:2005fu}. They turn out to produce a tiny
 disturbance to it, apart from avoiding level crossing. Hence, we 
expect the effects of
 $n_f\not=0$ to be important only when our states are very close to some heavy-light meson pair 
threshold. This is the reason why we quoted the location of nearby thresholds in ref. \cite{Oncala:2017hop}. %when identifying our hybrid candidates with XYZ states in section \ref{sec:exp}.
The second one is the appearance of light quark excitations, in addition to the gluon ones, in 
the 
static spectrum of Fig. \ref{kuti}. They may have different quantum numbers, for instance non zero isospin (in this
case they may be relevant to the experimentally discovered charged $Z$ states). %We do not know 
%anything about those and, 
As pointed out in \cite{Brambilla:2008zz} 
and more recently emphasized in \cite{Braaten:2013boa,Braaten:2014qka}, it would be extremely 
important to have lattice QCD evaluations of the static energies of light quark excitations (see \cite{Bicudo:2012qt} for work in this direction). We suspect that light quark excitations with the same quantum numbers as the gluonic ones will only provide small modifications 
to the hybrid potentials, since they correspond to higher dimensional operators. 
In this respect, it is significant that tetraquark models have also difficulties to encompass the $X(4140)$ in their spectrum together with $X(4237)$, $X(4500)$ and $X(4700)$ \cite{Esposito:2016noz}.% In fact, the $X(4140)$ structure may be due to a threshold enhancement according to some authors \cite{vanBeveren:2009dc,Liu:2016onn,Ortega:2016hde}.
%This means that tetraquarks with the same quantum numbers as hybrids will in general be hidden in the spectrum of the latter. 

\section{Conclusions}
We have calculated the charmonium and bottomonium hybrid spectrum in a QCD based approach, including for the first time the mixing with standard charmonium and bottomonium states. The latter leads to enhanced spin symmetry violations, which are instrumental to identify a number of XYZ states as hybrid states. Most of the isospin zero XYZ states fit well in our spectrum, either as hybrids or as standard quarkonium states. We have also estimated several decay widths and provided model independent formulas for the hyperfine splittings.
\label{sec:concl}

\section*{Acknowledgements}

Support from
the 2014-SGR-104 grant (Catalonia), the FPA2013-4657, FPA2013-43425-P, FPA2016-81114-P and FPA2016-76005-C2-1-P projects (Spain) is gratefully acknowledged.

\end{document}